\documentclass[reprint,prx,amsmath,amssymb,aps,superscriptaddress,nofootinbib,floatfix]{revtex4-2}
\usepackage{graphicx}
\usepackage{dcolumn}
\usepackage{bm}
\usepackage[utf8]{inputenc}
\usepackage{amsmath, amsthm, amsfonts, amssymb}
\usepackage[svgnames]{xcolor}
\usepackage[colorlinks=true,allcolors=NavyBlue]{hyperref}
\usepackage{physics}   
\usepackage{mathrsfs}

\newcommand{\xbb}{\bar{\bm x}}

\newcommand{\vbb}{\bar{\bm v}}

\newcommand{\diff}{\text{d}}

\begin{document}


\author{Carlos Barcel\'o}
\affiliation{Instituto de Astrof\'{\i}sica de Andaluc\'{\i}a (CSIC), Glorieta de la Astronom\'{\i}a, 18008 Granada, Spain}

\author{Valentin Boyanov}
\affiliation{Departamento de F\'{\i}sica Te\'orica and IPARCOS, Universidad Complutense 
de Madrid, 28040 Madrid, Spain}

\author{Luis J. Garay}
\affiliation{Departamento de F\'{\i}sica Te\'orica and IPARCOS, Universidad Complutense 
de Madrid, 28040 Madrid, Spain}

\author{Eduardo~Mart\'{\i}n-Mart\'{\i}nez}
\affiliation{Department of Applied Mathematics, University of Waterloo, Waterloo, Ontario, N2L 3G1, Canada}

\affiliation{Institute for Quantum Computing, University of Waterloo, Waterloo, Ontario, N2L 3G1, Canada}
\affiliation{Perimeter Institute for Theoretical Physics, 31 Caroline St N, Waterloo, Ontario, N2L 2Y5, Canada}

\author{Jose M. S\'anchez Vel\'azquez}
\affiliation{Departamento de F\'{\i}sica Te\'orica and IPARCOS, Universidad Complutense de Madrid, 28040 Madrid, Spain}
\affiliation{Instituto de F\'{\i}sica Te\'orica UAM-CSIC, Universidad Aut\'onoma de Madrid, Cantoblanco, 28049, Madrid, Spain}


\title{Warp Drive Aerodynamics}

\begin{abstract}
In this work we analyse the potential for a warp drive spacetime to develop instabilities due to the presence of quantum matter. Particularly, we look for points of infinite blueshift (which are analogous to points of a black hole inner horizon, known for its semiclassical instability), and categorise them through the behaviour of geodesics in their vicinity. We find that warp-drive bubbles in dimension 2+1 or higher are in fact likely to be stable, as they generally contain only isolated points where divergences are approached, leading to a finite limit for the overall accumulation of destabilising energy. Furthermore, any semiclassical instabilities in the warp drive due to energy-density buildups can be further diminished with particular, more ``aerodynamic" shapes and trajectories for the drive.
\end{abstract}

\maketitle

\section{Introduction}

Superluminal travel has been in humankind's collective imagination and popular culture for many decades. The possibility of visiting neighbouring stars, nebulae, or even galaxies and coming back, all within the lifespan of a single human being, is certainly appealing to any avid fan of science fiction. However, the launchpad for our future superluminal journeys is riddled with difficulties, which can be seen either as a  sign of unfeasiblity, or as an invitation to push forward technological boundaries and theoretical concepts.

From a physical standpoint, the fact that nothing can travel through space faster than light seemingly puts heavy restrictions on such journeys. However, general relativity circumvents this issue by allowing spacetime itself to ``move" in a way which can effectively increase or decrease the travel time between distant objects. There is no a priori limit on the velocity of this movement: the expansion and contraction of spacetime can even modify its causal structure. An example of this behaviour can be readily found in spacetimes with cosmological horizons, where objects recede away from each other faster than the speed of propagation of light. With this idea in mind, Alcubierre devised a spacetime geometry which combines a local expansion and contraction of spacetime to produce an apparent superluminal bubble: the warp drive~\cite{Alcubierre1994}. This geometry brings the idea of superluminal interstellar travel to the realm of physics. However, its construction requires the creation and manipulation of large quantities of exotic matter~\cite{anti-warp1,Alcubierre2017}. In other words, the stress-energy tensor which generates the warp drive solution of the Einstein equations violates every local energy positivity condition~\cite{HawkingEllis1973}. In fact, this is a manifestation of an even more general restriction: any asymptotically flat configuration which gives rise to superluminal travel appears to require exotic matter~\cite{Olum1998}.

As of yet, there is no experimental evidence of the existence of exotic matter capable of such spacetime distortions. In fact, attempts have been made to provide a geometric interpretation for the absence of gravitating exotic matter through the addition of an underlying causal structure which limits how classical spacetimes can curve~\cite{VisserBassett}. Alternatively, but in the same line of reasoning, this underlying causal structure could be less rigid and allow certain types of emergent warp-drive configurations, though never ones which produce closed timelike curves~\cite{BarceloGerardo2022}.

Other arguments suggest that exotic matter cannot be so easily dismissed, and can in fact be engineered at will. One comes from interpreting the Casimir effect in terms of quantum vacuum energy (for an alternative interpretation see~\cite{Jaffe2005}). Indeed, attempts have been made to construct the required negative energy profile present in a warp drive solution through manipulating the boundary conditions of quantum vacuum modes in a Casimir-effect manner \cite{White2021}. There are also other, more robust proposals for methods in which effective exotic matter distributions can be generated, such as the light-matter interaction and the protocol of quantum energy teleportation~\cite{NicoEdu}.

Additionally, effective exotic matter even appears naturally when the theory of quantum fields is formulated in curved spacetime backgrounds (a famous example of this being the Hawking evaporation of black holes \cite{Hawking1975}). However, given that in this theory it is the curvature of spacetime itself which makes quantum states react to produce these negative energies, it is not clear whether such effects would work in favour or against building configurations such as warp drives. In fact, a semiclassical analysis suggests the latter: assuming that we are free to manipulate exotic matter at will and can construct a warp drive, the simplest vacuum state defined on this curved spacetime has been argued to work against its stability~\cite{Hiscock1997,anti-warp2}.

Among all the issues mentioned above, this semiclassical instability is perhaps the most critical roadblock for the feasibility of warp-drive configurations of any size, seemingly banishing them forever to the realm of science fiction. In~\cite{Hiscock1997} it was established that a 1+1 dimensional warp-drive configuration, corresponding to the central axis of movement of higher-dimensional drives, develops this instability. Then, in \cite{anti-warp2} this analysis was generalised to warp drives formed dynamically from an initially flat spacetime. Calculating the renormalised stress-energy tensor (RSET) of a quantum scalar field, it was shown that the accumulation of geodesics, and correspondingly of modes of the quantum field, at the front end of the drive leads to an exponential growth in the vacuum energy density, analogous to that found at the inner horizon of black holes \cite{BalbinotPoisson93,Wald2019,Ori2019,Barcelo2021}. Furthermore, it was shown that this instability survives even in the presence of a modified dispersion relation at high energies~\cite{Coutantetal2012}. Thus, this instability, being caused by the very superluminal movement of the warp-drive with respect to the quantum vacuum, does indeed appear unavoidable.

At least, this is the case in 1+1 dimensions, but whether and to what degree this semiclassical instability is present in more realistic higher-dimensional warp-drive spacetimes, including in the 3+1 dimensions of our universe, has so far remained an open question. This is the problem we tackle in this work.

To give an exact measure of the energy associated with the potential instability, and thus the degree to which it would obstruct the prolonged existence of such geometries, we would need to calculate the RSET exactly; however, due to the technical difficulties involved in such a calculation in dimensions higher than 1+1 (and in the absence of certain symmetries), it is not the path taken in this work. Instead, we make use of some important intuitions gained from the 1+1 example, as well as from calculations in black hole spacetimes, to give a rough estimate of the vacuum energy in these warp drives of higher dimensions. Particularly, the exponentially growing accumulation of energy can be related to the presence of surfaces (or points) of infinite blueshift, which are known to cause instabilities even on a classical level (cf. mass inflation instability of black holes with an inner horizon \cite{Poisson1989}). Therefore, a classical analysis of geodesics can likely suffice to identify the regions which may cause such instabilities.

In this work we analyse the geodesics of a warp-drive spacetime in 2+1 dimensions, focusing in particular on the vicinity of the walls of the warp bubble, which posses horizon-like properties. Remarkably, we find that for warp bubbles of finite spatial extension there is generally only a single point where infinite blueshift can occur, suggesting that the semiclassical singularity in 2+1 and higher dimensions is far weaker than its 1+1 dimensional counterpart. Particularly, by looking at the geodesics trapped in an approach toward this point, as well as ones which get close to it but end up deflected away, we estimate that the integrated semiclassical energy density around this point should be finite in most cases. Furthermore, we show that although changing the shape and trajectory of the warp bubble cannot eliminate this point, it can serve to further disperse the geodesics in its vicinity and, by extension, the semiclassical energy accumulation. Much like how aircraft reduce their air resistance by having particular shapes and adapting to air currents, warp drives must adapt their geometry and dynamics in order for the quantum vacuum to offer as little resistance as possible to their movement.

In section \ref{s2} we provide a brief introduction to the warp drive and its semiclassical instability in 1+1 dimensions. In section \ref{s3} we analyse the geodesics in a 2+1 dimensional drive which are relevant for determining its causal structure and its potential instability-inducing points. We estimate the semiclassical energy accumulation at and around these points, and we analyse how this accumulation may be dispersed by changing the shape or trajectory of the warp bubble. In section \ref{s4} we provide some concluding remarks.

\section{Warp drive and the semiclassical instability}\label{s2}

The Alcubierre warp drive as an isolated system in an asymptotically flat spacetime has the following metric:
\begin{equation}\label{metricx}
\text{d}s^2=-c^2\text{d}t^2+[\text{d}{\xbb}-{\vbb}(t,{\xbb})\text{d}t]^2,
\end{equation}
where ${\xbb}$ represents spatial coordinates and ${\vbb}(t,{\xbb})$ determines the velocity and shape of the warp bubble (both these quantities are defined as Euclidean vectors with as many components as spatial dimensions in the manifold). Flat spacetime is recovered far away from the bubble by imposing that $|{\vbb}|\to 0$ as $|{\xbb}|\to\infty$. We take $\bar{\bm x}_{\text{c}}(t)$ to be the trajectory of the centre of the bubble in this asymptotically-Minkowskian coordinate system, and for convenience we also define the comoving spatial coordinates ${\bm x}={\xbb}-\bar{\bm x}_{\text{c}}(t)$. We can then write the usual definition ${\vbb}(t,{\bm x})=\bar{f}({\bm x}){\bm V}(t)$, where ${\bm V}(t)=\diff\bar{\bm x}_{\text{c}}(t)/\diff t$ is the velocity of the bubble and $\bar{f}({\bm x})$ determines its shape. At the centre of the bubble this shape function must satisfy $\bar{f}(0)=1$, and as $|{\bm x}|\to\infty$ it must tend to zero sufficiently quickly. In comoving coordinates the metric can be written as
\begin{equation}\label{metricr}
\text{d}s^2=-c^2\text{d}t^2+[\text{d}{\bm x}+{\bm v}(t,{\bm x})\text{d}t]^2,
\end{equation}
where ${\bm v}(t,{\bm x})=f({\bm x}){\bm V}(t)$, with $f=1-\bar{f}$.

\begin{figure}
    \centering
    \includegraphics[scale=0.55]{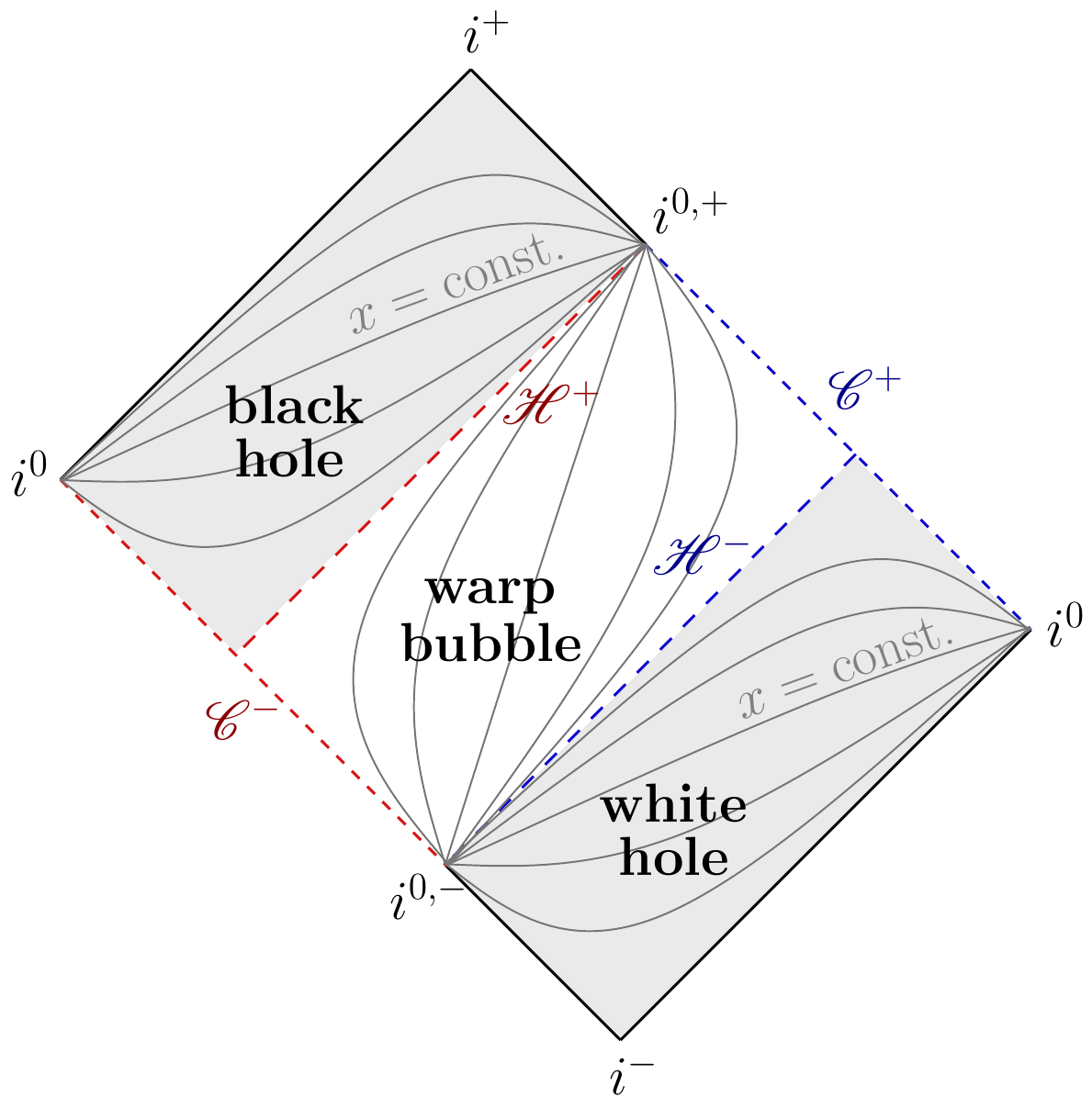}
    \caption{Causal structure of a stationary 1+1 dimensional warp drive spacetime. The warp bubble is located between a white and a black hole, separated by the past and future horizons, denoted by $\mathscr{H}^-$ and $\mathscr{H}^+$ respectively. $\mathscr{H}^+$ connects with a Cauchy horizon to the past $\mathscr{C}^-$, and $\mathscr{H}^-$ with a Cauchy horizon to the future $\mathscr{C}^+$. The curved lines correspond to lines of $x=\text{const.}$, which are timelike inside the warp bubble, and spacelike everywhere else.}
    \label{f1}
\end{figure}

To understand this geometry better, let us start with a 1+1 dimensional stationary (${\bm V}(t)=\text{const.}$) case. In this case, the line element acquires the same form as that of the radial-temporal sector of a black hole spacetime written in Painlevé-Gullstrand coordinates~\cite{Martel2001}. Particularly, the front end of the warp drive behaves like a white hole horizon (more precisely, a white hole outer horizon or, equivalently, a black hole inner horizon), and the rear end like a black hole horizon (time reverse of the former). In comoving coordinates, the inside of the warp bubble appears located between two trapped regions, i.e. between a white and a black hole, as shown in Fig.~\ref{f1}.

In \cite{anti-warp2} it was shown that when such a configuration forms from an initially flat region, the presence of any background quantum field, even in vacuum,\footnote{The instability studied here is the same for any Hadamard state.} leads to an exponential growth of energy at the white hole horizon. Particularly, the energy density obtained when contracting the RSET $\expval{T_{\mu\nu}}$ with the velocity $u^\mu$ of a free-falling observer which approaches this horizon grows as
\begin{equation}\label{2div}
\rho=\expval{T_{\mu\nu}}u^\mu u^\nu\sim e^{2\kappa t},
\end{equation}
where $\kappa$ is the surface gravity of the horizon. Around this horizon, the spacetime exhibits the same causal features as a black hole with an inner horizon, developing a Cauchy horizon and a corresponding blueshift instability. Semiclassically, it is precisely the blueshift  which causes the behaviour observed in \eqref{2div}, as has been seen in charged and rotating black holes \cite{BalbinotPoisson93,Wald2019,Ori2019,Barcelo2021,Zilberman2022}.
How this instability behaves in higher-dimensional configurations therefore depends entirely on how the horizon structure shown in Fig. \ref{f1} generalises to them. As it happens, for generic, finite-sized warp bubbles, the dimension of the surface of infinite blueshift does not grow with the dimension of the spacetime, but remains the same as in 1+1. We will now proceed to show this explicitly in 2+1 dimensions.

\section{Geodesics and stability in 2+1 dimensions}\label{s3}

Let us now turn our attention to the 2+1 dimensional warp drive. We will start with a thorough analysis of a particularly simple, yet quite generic configuration: a stationary warp bubble travelling in a straight line, with a geometry which has a reflection symmetry with respect to a central axis aligned with the direction of movement, as depicted in Fig. \ref{f2}. The comoving spatial coordinates will be denoted by $\{x,y\}$, where $x$ is taken to be aligned with the direction of motion and $y$ with the direction of symmetry. The line element of the geometry is
\begin{equation}\label{metric2}
\diff s^2=-\diff t^2+[\diff x+v(x,y)\diff t]^2+\diff y^2.
\end{equation}
Taking $y=0$ as the position of the central axis, the function $v(x,y)$ has even parity in $y$. The equations which determine the null geodesics of this spacetime are
\begin{align}
(v^2-1)\dot{t}^2+2v\,\dot{t}\dot{x}+\dot{x}^2+\dot{y}^2&=0,\label{geo0}\\
(v^2-1)\,\dot{t}+v\,\dot{x}&=E,\label{geo1}\\
\ddot{y}-\partial_y v(v\,\dot{t}^2+\dot{t}\dot{x})&=0\label{geo3},
\end{align}
where $E$ is an integration constant and the dot indicates differentiation with respect to the geodesic affine parameter $\sigma$.

\subsection{Movement on the central axis}

\begin{figure}
    \centering
    \includegraphics[scale=0.8]{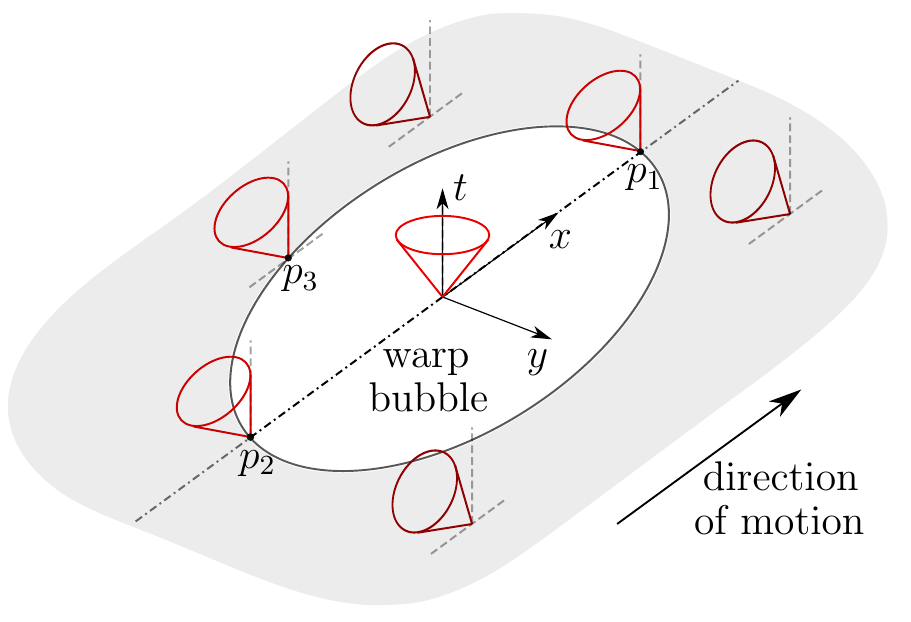}
    \caption{Warp bubble in 2+1 dimensions as seen in comoving coordinates. Light cones show the permitted directions of movement for causal trajectories. The point $p_1$ is the front end of the bubble, which produces a blueshift instability toward the future. Likewise, $p_2$ produces an instability toward the past. $p_3$ represents a generic point of the border of the bubble which does not lie on the central axis.}
    \label{f2}
\end{figure}

The 1+1 dimensional example presented above corresponds to the movement of geodesics along the central axis. We can recover the causal structure of Fig.~\ref{f1} by integrating the null geodesic equations with the initial conditions $y_0=\dot{y}_0=0$ (which, through Eq. \eqref{geo3} and the fact that $\partial_yv=0$ at $y=0$, implies $y(\sigma)=0$). It is worthwhile to do this explicitly in the vicinity of the edges of the bubble, where $v$ approaches 1. Let $p_1=(x_1,0)$ be the front end of the bubble, as shown in Fig. \ref{f2}. Near this point we can consider the series expansion in the $x$ direction
\begin{equation}\label{expan1}
v(x,0)=1+\kappa (x-x_1)+\cdots,
\end{equation}
where $\kappa$ is a positive constant. At leading order, equations \eqref{geo0} and \eqref{geo1}, with the expansion \eqref{expan1}, have a family of solutions
\begin{equation}\label{csol}
x-x_1\simeq-E(\sigma-\sigma_1),\qquad
t\simeq-\frac{1}{\kappa}\log|\sigma-\sigma_1|,
\end{equation}
with $\sigma_1$ a constant. We see that these null geodesics reach the limit $t\to\infty$ at a finite $\sigma$, corresponding to the future Cauchy horizon in Fig. \ref{f1}. The past Cauchy horizon is obtained analogously by taking the expansion \eqref{expan1} at the rear end of the drive ($p_2$ in Fig. \ref{f2}), making $\kappa$ negative, and considering ${t\to-\infty}$.

We also note that, given that \eqref{geo0} is quadratic in the coordinate functions, there are in fact two families of solutions of the geodesic equations around $p_1$. Aside from the ones shown above, there are also solutions for which $t$ does not diverge, which can be identified as the ones which cross the black and white hole horizons depicted in Fig. \ref{f1} (left-moving light rays). Locally, these can be seen as backwards-directed trajectories relative to the bubble.

The blueshift instability is triggered by the rays which take infinite $t$ to reach $p_1$, as can be seen from the fact that the spatial separation between any two distinct lightlike observers which approach this point tends to zero, implying that the wavelength of individual perturbations also goes to zero, i.e. they are infinitely blueshifted. Intuitively, one can then see that backreaction from generic perturbations may well destabilise this configuration even on a classical level. The semiclassical argument follows the same line of reasoning, though it can lead to a stronger singular behaviour, as is the case in black holes \cite{Wald2019}.

\subsection{Other unstable points?}

We have shown that the causal structure of the 1+1 case continues to be present in higher dimensions, at least on one axis. However, it is not clear whether other points of infinite blueshift besides $p_1$ (and $p_2$, if we consider past instabilities) are present in other parts of the 2+1 configuration. In fact, for the simplest type of warp bubble, it turns out that there are no other such points, as we will now show.

The points which one can expect to have special causal behaviour are the ones which comprise the rest of the edge of the warp bubble, where $v=1$. Let $p_3$ be one such point (see Fig. \ref{f2}). At this point, we can use the labels $\kappa=\partial_xv|_{p_3}$ and $\zeta=\partial_yv|_{p_3}$. We define the front and back end of the bubble ($p_1$ and $p_2$ in Fig. \ref{f2}) as the points where $\zeta=0$, which for the symmetric bubble we are considering lie on its intersection with the symmetry axis $y=0$. These are the points which correspond to the future and past Cauchy horizons shown above. For a smooth convex bubble, $\zeta\neq 0$ at all other points of its frontier.

We are interested in whether there are geodesics for which $t$ diverges at finite $\sigma$ when approaching $p_3$. We can answer this by substituting $v$ and its derivatives for their values on these points in the geodesic equations, and checking whether $t$ can approach infinity while $\sigma$, $x$, and $y$ remain bounded. In Eq. \eqref{geo1}, the first term on the left-hand side, $(v^2-1)\dot{t}$, can tend either to 0, a constant, or infinity. If it went to infinity, then $\dot{x}$ would also diverge at $p_3$, which, given the analiticity of the geometry, leads to no consistent solutions (one can check this explicitly by taking an arbitrary inverse-polynomial or logarithmic divergence for $\dot{x}$ and checking the requirements imposed on the other derivatives in equations \eqref{geo0} and \eqref{geo3} at $p_3$, arriving at an inconsistency). If this first term of \eqref{geo1} goes to a constant or to 0, then equations \eqref{geo1} and \eqref{geo3} become
\begin{equation}
\dot{x}|_{p_3}=\tilde{E}=\text{const.},\qquad \ddot{y}|_{p_3}=\zeta(\dot{t}^2+\tilde{E}\dot{t})|_{p_3}.
\end{equation}
Since the geometry is analytic, the divergence of $t$ implies the divergence of its derivatives. Therefore, $\dot{t}$ and $\dot{t}^2$ would have different rates of divergence, and $\ddot{y}$ would remain finite only if $\zeta=0$. If $\zeta\neq 0$, then $\ddot{y}$ diverges, which does not occur for any consistent solutions at the finite point $p_3$ (this can again be seen by considering the analiticity of the geometry or checking explicitly for such solutions in the geodesic equations).

In other words, the only points on the boundary of the bubble $v=1$ which have the possibility of generating Cauchy horizons are the ones where $\partial_yv=\zeta=0$, i.e. where the derivative of the shape function $v$ in the direction perpendicular to that of motion is zero. Stated as such, this result can be seen to be independent of the particular shape or symmetry of the bubble, as long as the configuration is stationary. For a smooth and convex bubble, this implies that there are strictly only two points of infinite blueshift akin to the ones present in 1+1 dimensions, and time symmetry tells us that only one is unstable toward the future (and the other toward the past).

If the bubble is not convex and additional points of $\partial_y v=0$ are present, then the warp drive could be said to be less ``aerodynamic" in its motion within the quantum vacuum, as it would find further resistance to its stability due to larger energy accumulation. However, as long as such points are isolated from each other, the overall configuration could potentially be stable, as the total amount of accumulated vacuum energy could be finite.

\subsection{Vicinity of the unstable points and vacuum energy divergence}

To find out whether the single-point blueshift instabilities present in 2+1 (and higher) dimensions are actually detrimental to the stability of the whole warp drive configuration, we must estimate the behaviour of the quantum vacuum energy in a small vicinity around these points. An instability is present only when the divergence at these points has a certain ``width", enough to produce a singularity if backreaction is considered. Finding out whether this is the case would generally involve calculating the RSET on this spacetime for a test field in an appropriate vacuum state. However, due to the great technical difficulty involved in such a calculation, we will resort to an estimation based on an extension of the analogy between the movement of geodesics at and around the central axis, and the 1+1 dimensional case.

To set this up, let us begin by considering a solution slightly away from the central axis solutions \eqref{csol}, but still in the vicinity of $p_1$. We now perform the expansion of $v$ around $p_1$ to leading order in both $x$ and $y$,
\begin{equation}\label{shapey}
v(x,y)\simeq 1+\kappa(x-x_1)+\xi y^{2n},
\end{equation}
where $\xi$ is a constant with appropriate inverse-length dimensions, positive if the bubble is convex and negative if it has a concave peak (and zero if it has a finite-sized flat peak), and $n$ a natural number. Larger values of $n$ make the peak more flat in the $y$ direction. We consider the deviation from the solution \eqref{csol} (where $y=0$),
\begin{align}
\delta x(t)&=x(t)-x_1-Ee^{-\kappa t},\label{delx}\\
\delta t(\sigma)&=t(\sigma)+\frac{1}{\kappa}\log|\sigma|,
\end{align}
where $E$ is a constant. It is convenient to rewrite equations \eqref{geo0}, \eqref{geo1} and \eqref{geo3} at leading order in $\delta x$, $y$ and $\delta t$ (and their derivatives) as
\begin{align}
y''+\kappa y'-2n\xi y^{2n-1}&\simeq 0,\label{devy}\\
\delta x'+\kappa\delta x+\xi y^{2n}+\frac{1}{2}(y')^2&\simeq 0,\label{xprime}\\
\delta\dot{t}+\frac{e^{2\kappa t}}{2\kappa^2 E}(2\kappa\delta x+2\xi y^{2n}+\delta x')\label{dott}&\simeq 0,
\end{align}
where the prime indicates a derivative with respect to~$t$. Eq.~\eqref{devy} can give us a description of the perturbation in $y$, from where we can use Eq. \eqref{xprime} to obtain the perturbation $\delta x$, and Eq. \eqref{dott} to find the modification $\delta t$ to $t(\sigma)$. Particularly, Eq. \eqref{devy} can be solved directly, and the validity of the solution can be checked by making sure that the approximations which lead to \eqref{devy} are accurate, which can be done with the solutions of \eqref{xprime} and \eqref{dott}.

Let us begin by looking at the case of a warp bubble edge with a finite-sized region which is flat in $y$. This, as one might imagine, is not a very ``aerodynamic" shape, as it is not convex. The solutions of \eqref{devy} with $n=0$ tending to this frontal region would be
\begin{equation}\label{flat}
y\simeq c_2+c_3e^{-\kappa t},
\end{equation}
where $c_{2,3}$ are integration constants. The constant $c_2$ is indicative of the fact that $p_1$ is no longer the only point which traps geodesics into a tendency toward a Cauchy horizon. Eq. \eqref{xprime} with $n=0$ furthermore shows us that for these solutions $x$ has the exact same behaviour at large $t$ as it does on the central axis (i.e. $\delta x$ has the same solutions as $x$ in \eqref{csol} and can be absorbed in the integration constants of the latter); and Eq. \eqref{dott} shows the same for $t(\sigma)$, implying that the approximations used to obtain \eqref{devy} are accurate. Therefore, in this case there would be a finite-sized region with a blueshift instability, and we can expect that this configuration would be unstable under both classical and semiclassical perturbations.

If the bubble has, say, a parabolic profile in $y$ (i.e. $n=1$), then the solutions become
\begin{equation}\label{sol}
y\simeq c_2e^{\eta_-t}+c_3e^{-\eta_+t},\qquad
\delta x\simeq c_1\,y^2,
\end{equation}
with $c_1$, $c_2$ and $c_3$ constants, and
\begin{equation}
\eta_\pm=\frac{\kappa}{2}\left(\sqrt{1+8\frac{\xi}{\kappa^2}}\pm 1\right).
\end{equation}
Let us first look at the case of a convex bubble, for which $\xi>0$ and consequently $\eta_\pm>0$. In this case, only initial conditions which give $c_2=0$ correspond to geodesics trapped in a tendency toward the tip of the bubble from outside the axis, since $y\to 0$ as $t$ grows. These fine-tuned geodesics (of measure zero within the total set of solutions) for each value of $c_3$ represent a separatrix between solutions deflected away (exponentially quickly, while the approximation is valid) to one side ($c_2>0$) and the other ($c_2<0$). As one might expect, the approximations leading to Eq. \eqref{devy} break down quickly when $c_2\neq0$, and become asymptotically exact when $c_2=0$.

\begin{figure}
    \centering
    \includegraphics[scale=0.5]{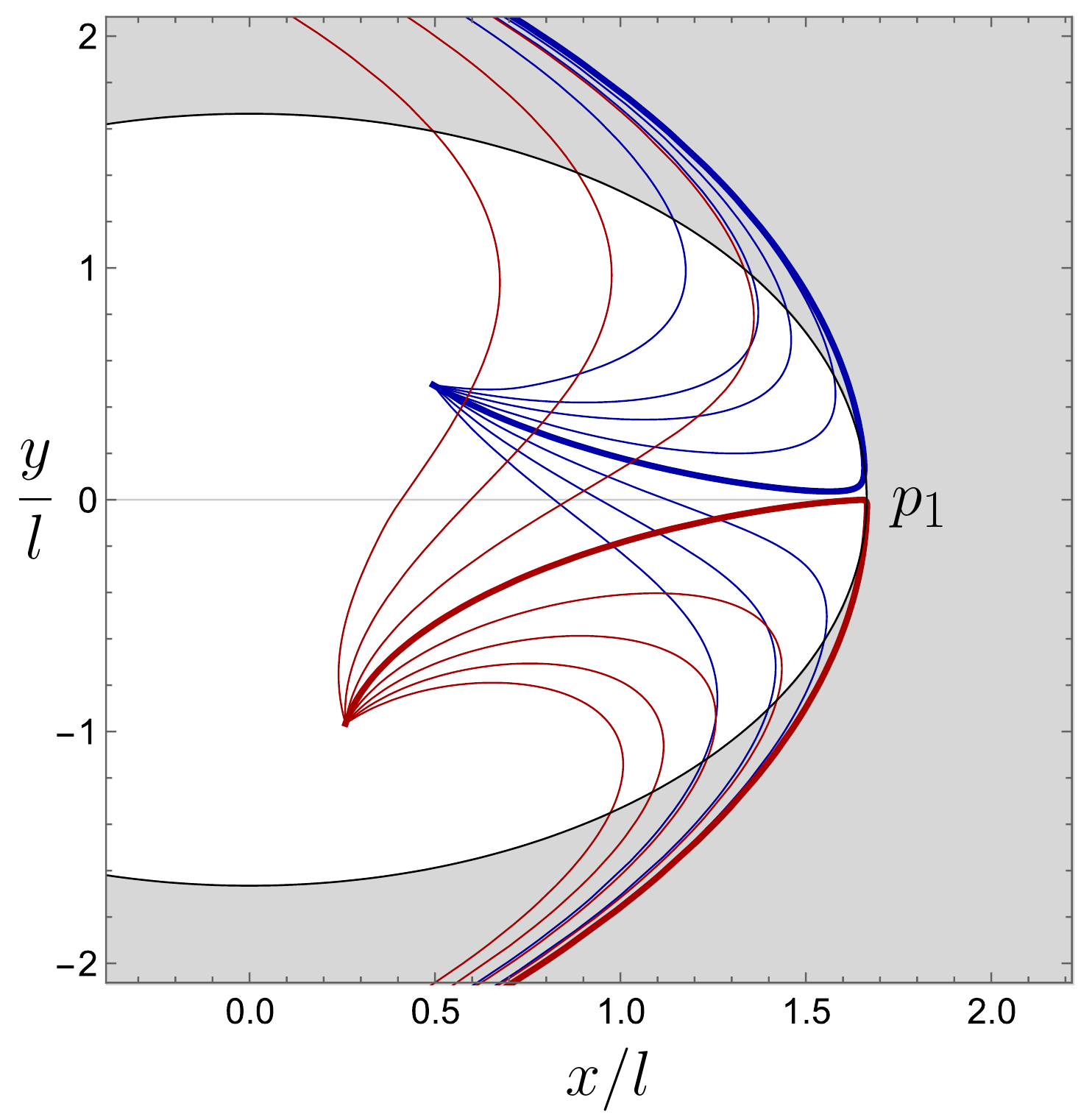}
    \caption{Numerical integration in time of two bundles of null geodesics launched from two different points in the interior of a stationary circular warp bubble moving at twice the speed of light. $l$ is a characteristic length scale of the bubble. The geodesics are launched in the forward direction (with respect to the motion of the drive), with an initial angle dispersion of $\pi/3$ between the first and last of each bundle of curves. The thicker lines of each bundle represent the geodesics which get closest to the point $p_1$, which in a vicinity of this point correspond to curves \eqref{sol} with small values of $c_2$.}
    \label{f3}
\end{figure}

A further check of this behaviour was performed numerically by directly solving equations \eqref{geo0}, \eqref{geo1}, and \eqref{geo3} for geodesics launched from within a moving convex warp bubble. The result is shown in Figure \ref{f3}, where one observes clearly the general behaviour of light within this geometry. The first thing we note about the geodesics shown is that they follow the restrictions imposed by the light cones represented in Fig. \ref{f2}: they can only move forward (toward larger values of $x$) while inside the bubble, and when they approach its edges they turn around. The thicker lines of each bundle of geodesics mark two curves which get close to the central axis in the vicinity of the point of runaway blueshift $p_1$. In these curves we see explicitly the behaviour captured in Eq. \eqref{sol} for solutions with small values of $c_2$. They initially approach the axis, until the exponential growth of $e^{\eta_-t}$ overcomes the smallness of $c_2$ and pushes them away. As expected, for geodesics launched from each point, the separatrix ($c_2=0$) between the ones which end up on the left and on the right of $p_1$ turns out to be impossible to capture numerically. This provides further evidence of the fact that, although these solutions end up infinitely blueshifted, they are of measure zero within the whole family of geodesics.

For a bubble with a more flat profile in $y$ at $p_1$ (i.e. $n>1$), there are no analytical solutions to \eqref{devy}, but it can be seen that the source term for the derivatives is smaller and deflection therefore has an initially polynomial (rather than exponential) dependence on $t$. Aside from this, the qualitative behaviour of the geodesics in such a bubble remains the same (this has been checked numerically).

Returning to the $n=1$ parabolic profile, we can make an important observation regarding the deflected geodesics. By taking the geodesic from \eqref{sol} with $c_3=0$ as representative of the generic qualitative behaviour of geodesics with $c_2\neq 0$, we can write its solution in terms of the initial condition $y(0)=y_0$ as
\begin{equation}
y\simeq y_0e^{\eta_-t}.
\end{equation}
We can then define a deflection time $t_{\rm def}$ as the time it takes for the solution to reach a fixed reference point $y_{\rm def}$,
\begin{equation}\label{tdef}
t_{\rm def}=\eta_-^{-1}\log(y_{\rm def}/y_0).
\end{equation}
The value of $y_{\rm def}$ is a characteristic length scale of the geometry which can be defined e.g. as the separation for which the approximation which lead to \eqref{devy} fails. The important part is the dependence on $y_0$, particularly, the logarithmic divergence as $y_0\to 0$.

In 1+1 dimensions, the presence of a Cauchy horizon and the corresponding divergence of $t$ for finite $\sigma$ in \eqref{csol} drives the exponential growth of the energy density \eqref{2div}. In 2+1 and higher dimensions, one may then expect the same type of growth only around points where the null geodesics behave the same way (i.e. as if approaching a Cauchy horizon), which occurs only when they approach the tip of the warp bubble. In other words, the exponential growth of the energy density should only occur at a single point. As to what happens in the vicinity of this point and how this energy accumulation falls off away from it, the geodesics with $c_2\neq 0$ might provide a clue.

Particularly, in the regime $t\ll t_{\rm def}$ these geodesics have a very small deviation from the central solution \eqref{csol}, and one might expect that they bring about a growth similar to \eqref{2div}, but instead of blowing up to infinity as $t$ grows, tending to a finite cutoff value with a profile given by the logarithm of $y_{\rm def}/y_0$ in \eqref{tdef}, with $y_0$ representing the separation from the central point.

Another argument in favour of this kind of asymptotic density profile can be made by just considering the blueshift of light rays which could be randomly launched in the general direction of the front of the warp bubble. The ones which happen to tend exactly to the tip are the only ones which are trapped and have a divergent tendency in their blueshift. In the rest of the bubble, one may expect that a stationary situation is quickly reached if e.g. the rays are launched at regular intervals. In a given time the same number of rays would enter a given area as the ones which exit it, though the closer this area is to the tip of the bubble, the longer their stay there and the larger their transient blueshift, giving rise to the same logarithmic profile of energy density.

In dimensions larger than 1+1, the ``instability" for a convex warp bubble is therefore just the growth at a single point, and integrating the energy in its vicinity the result would not asymptotically tend to a divergence. Even if the logarithmic profile we obtain for the asymptotic tendency for the semiclassical energy density is not the correct description one would get from calculating the RSET exactly in 2+1 or higher dimensions, at the very least the fact that this profile is related to the accumulation of geodesics is robust. Therefore, additional dispersion of these geodesics would translate into further stabilisation of the semiclassical behaviour, as can be achieved by decreasing $t_{\rm def}$ (e.g. by making the peak of the bubble sharper in $y$, i.e. making $\xi$, and hence $\eta_-$, larger), or by making the trajectory of the drive deviate from the straight line path we have considered here, as we will show numerically below. On the other hand, making $t_{\rm def}$ larger (e.g. by decreasing $\xi$ or increasing $n$) would have the opposite effect and bring the drive closer to instability.

A convex shape with a very sharp peak, which offers the least resistance for travel in the presence of a quantum field, even in vacuum, is reminiscent of the shapes used for supersonic aircraft which minimise the frontal pressure and drag that they experience. By extension of this analogy, one would naturally expect that a warp bubble with a flat or concave peak would experience much more resistance, i.e. a much stronger blueshift instability. Indeed, in the case of a flat peak we found that the solutions which are trapped in a tendency toward a Cauchy horizon are much more abundant \eqref{flat}. For a concave peak with e.g. a locally parabolic profile, the solutions would be the same as \eqref{sol} but, $\xi$ being negative, $\eta_-$ would have a negative real part, making both exponentials decreasing ones. This would be a case in which a divergence of the order of that of a flat peak is concentrated at a single point, making the instability even greater.

\subsection{Numerical analysis of non-stationary configurations: further stabilising the warp drive}

In light of these results, one may wonder how this behaviour generalises to dynamical warp drive spacetimes, i.e. ones in which the warp bubble can change its trajectory and velocity over time. Particularly, we want to see whether the point of divergent blueshift $p_1$ remains, or whether some trajectories for the bubble can ``shake off" the potentially accumulated geodesics around such a point at regular intervals.

There are two types of movement which have the potential to do this: a change in direction, or a temporary reduction of the velocity to a subluminal one. However, we have found through a numerical analysis that neither one of these can fully eliminate the point of asymptotically infinite blueshift (and its corresponding finite-time accumulation of vacuum energy). Nonetheless, they can significantly disperse the geodesics in its vicinity, producing the same effect as making the peak of a straight-line bubble sharper.

\begin{figure}
    \centering
    \includegraphics[scale=0.55]{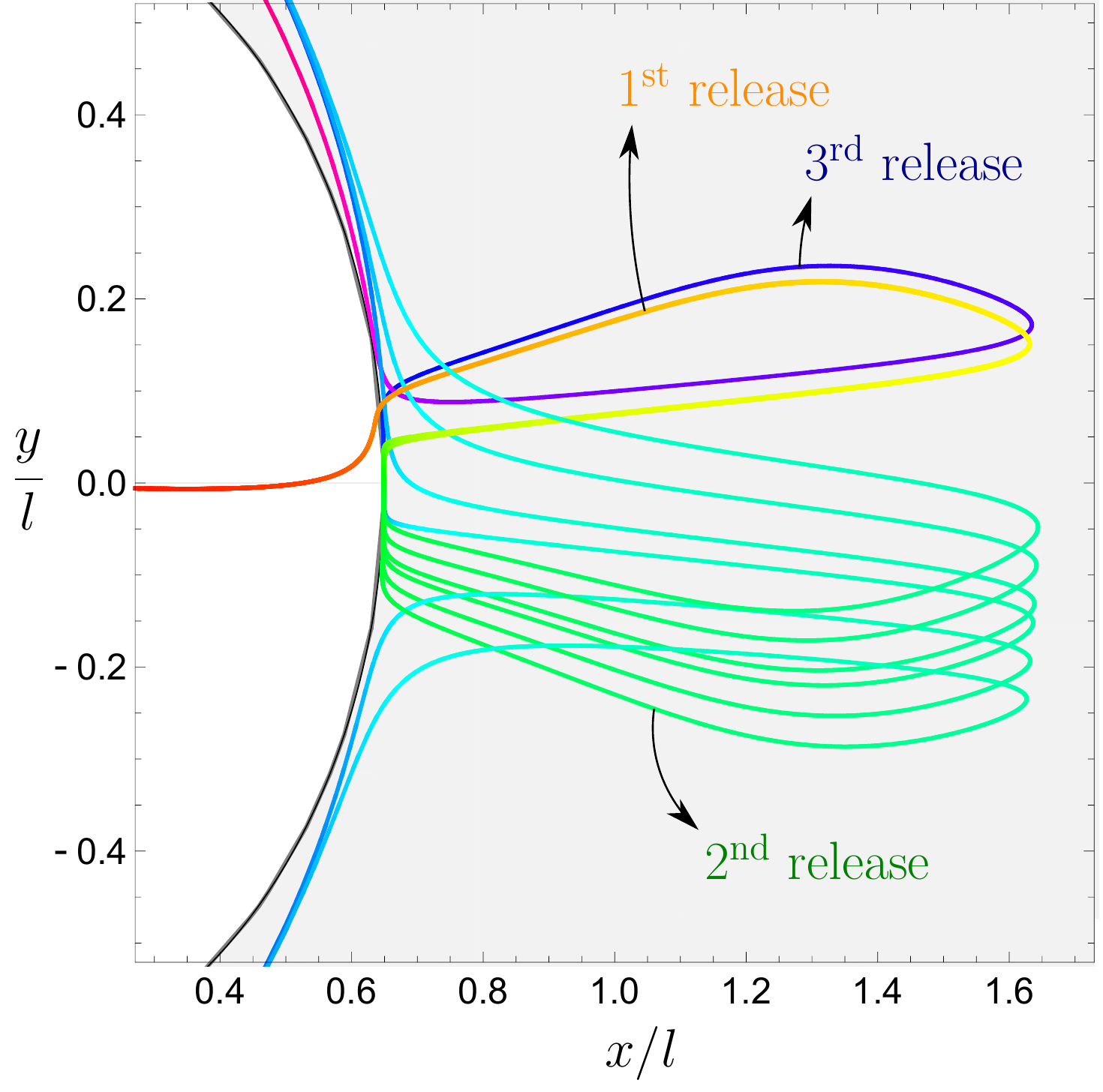}
    \caption{Numerical integration of null geodesics launched from the centre of a circular warp bubble moving in a zig-zag pattern with subluminal and superluminal intervals. The velocity of the drive in the $x$ direction changes between 0.2$c$ and 10$c$ with frequency $2\pi/5$, while in the $y$ direction it changes between 0.8$c$ and -0.8$c$ with frequency $\pi/5$. There are 6 geodesics emerging from the left side of the plot. The angle at which they are emitted only varies by $10^{-4}$ between them, so they initially overlap. Each time the drive becomes subluminal, the geodesics are released forward, only to turn around once it becomes superluminal again. At the second release they become dispersed enough to be visibly separate, while at the third release only one of them remains (the rest having been dispersed to the sides).}
    \label{f4}
\end{figure}

The fact that the equivalent of $p_1$ cannot be eliminated can be deduced from a simple consideration: if the warp bubble continues to exist indefinitely (and has a well-defined asymptotic direction), for geodesics launched from each point in its interior there will always be a separatrix between those which end up on one side or the other of its asymptotic trajectory. What can be controlled, however, is the amount of blueshift this separatrix and its adjacent geodesics experience at finite times.

As an example, we present the result from a numerical analysis of a dynamical warp drive configuration which combines the two modifications to the straight-line scenario mentioned above: a change of direction, achieved through a periodically varying velocity in the $y$ direction, and a change to a subluminal velocity in $x$, also performed periodically. Figure \ref{f4} represents the key features of the behaviour of light rays in such a geometry. Most null geodesics launched from inside the bubble quickly escape to either side of the averaged direction of motion, and never approach the bubble again, akin to those plotted in Fig. \ref{f3}. The main difference is seen in the rays which are launched approximately in the direction of motion. Particularly, those which remain in the vicinity of the front end of the bubble long enough to catch one of the changes in velocity have the chance to move in the forward direction beyond the confines of the bubble while the drive is subluminal. Then, when the drive becomes superluminal again, the bubble catches up to those rays once again, and they are now deflected to either side of it from the outside. Those which are close to the separatrix between the ones deflected to either side can again remain in the vicinity of the edge of the bubble long enough to catch the next change in velocity and move forward again. This process repeats periodically, and there is once again a particular set of trajectories (which are of measure zero within the total set of null geodesics, and which asymptotically coincide) that define the separatrix between rays deflected to either side of the drive.

Figure \ref{f4} represents 6 null geodesics which are launched in an approximately forward direction, with an initial angle dispersion of the order of $10^{-4}$ (making them overlap initially, on the left side of the plot). We see that each time the drive becomes subluminal, the geodesics are allowed to leave the bubble in the forward direction (or, as seen from inside the drive, it is the bubble that effectively expands to infinity). Then when it picks up superluminal speed they again turn around. Due to the oscillatory nature of the movement, we can expect that the separatrix also describes a periodic movement in space. In fact, one of the geodesics in Fig. \ref{f4} is very close to such a behaviour: after the third time it is released in a forward direction (i.e. the third time the bubble becomes subluminal) it nearly follows the same trajectory as after the first time, though the small difference makes it so it is deflected away (toward positive $y$) in the end.

While a convoluted trajectory for the warp drive would have a negative impact on its initial purpose (i.e. shortening travel time), it can, on the other hand, increase its semiclassical stability by reducing the accumulated blueshift around its peak. Constructing an optimal warp bubble shape and trajectory would therefore become a balancing act between having a short travel time and minimising the (possibly already very small) accumulation of unwanted vacuum energy and blueshifted classical perturbations. Of course, this problem would likely have a secondary role when compared to the inevitable engineering difficulties in constructing such configurations to begin with.

\section{Conclusions}\label{s4}

We have analysed the semiclassical instability present in the Alcubierre warp-drive spacetime through its relation to the behaviour of null geodesics. We have argued that the strong instability found in 1+1 dimensional configurations can actually be tamed in spacetimes of higher dimensions by choosing appropriately the shape and the trajectory of the warp bubble.

First, the warp field should be chosen to have an ``aerodynamic" shape, so as to deflect null geodesics away from its unstable point in the shortest time possible. Second, the trajectory of the drive can be chosen so as to further facilitate this dispersion, particularly with slight changes in its direction of movement (e.g. a small ziz-zag component to the motion), and with alternating intervals of subluminal and superluminal warp field velocities.

Our findings are interesting even from a purely geometrical perspective. In 1+1 dimensions the front wall of a warp drive acts as a pure inner horizon. However, in higher dimensions the warp drive does not have a closed inner horizon (or indeed any closed trapped surfaces). Instead, the warp-drive bubble can be interpreted as an interpolation between an inner horizon point (the front end of the bubble) and an outer horizon point (the back end of the bubble); then, in between we have a causal structure more similar to that of an ergoregion, from which signals can in fact escape. This is the reason why the geodesic accumulation, and the corresponding blueshift instability, is limited to only single points, at least when the shape of the bubble is smooth.

From a more physical perspective, while this work does not prove that warp drives are a completely viable option for faster-than-light travel, we do present strong evidence that the semiclassical instabilities do not, in principle, preclude ``aerodynamic" configurations of the Alcubierre drive from being able to sustain a net superluminal speed. The launchpad for our future superluminal journeys is thus left with one less obstruction.

\begin{acknowledgments}
 E. M-M. is funded by the NSERC Discovery program as well as his Ontario Early Researcher Award. Financial support was provided by the Spanish Government through the projects PID2020-118159GB-C43, PID2020-118159GB-C44 (with FEDER contribution), and by the Junta de Andaluc\'{\i}a through the project FQM219. C. B. acknowledges financial support from the State Agency for Research of the Spanish MCIU through the ``Center of Excellence Severo Ochoa" award to the Instituto de Astrof\'{\i}sica de Andaluc\'{\i}a (SEV-2017-0709). V. B. is funded by the Spanish Government fellowship FPU17/04471.
\end{acknowledgments}

\bibliography{biblio_def}

\end{document}